\documentclass[letterpaper,twocolumn,10pt]{article}

\usepackage{usenix-2020-09}

\makeatletter
\def\@maketitle{\newpage
  \vbox{
    \vspace*{22pt}             %
    \centering
    {\Large\bfseries \@title\par}
    \vskip 12pt               %
    {\large\itshape
      \lineskip 0pt
      \begin{tabular}[t]{c}
        \@author
      \end{tabular}\par
    }
    \vskip 13pt                %
  }
}
\makeatother

\AtBeginDocument{%
  }

\usepackage{subcaption}
\usepackage{graphicx}
\usepackage{amsmath}
\usepackage{hyperref}
\usepackage{cleveref}
\crefname{section}{§}{§§}
\Crefname{section}{§}{§§}
\usepackage{algorithmicx}
\usepackage{algorithm}
\usepackage{algpseudocode}
\usepackage[utf8]{inputenc}
\usepackage{bbding}
\usepackage{pifont}
\usepackage{multicol}
\usepackage{multirow}
\usepackage{makecell}
\usepackage{xspace}
\newcommand{\Sys}[0]{ETH\xspace}
\usepackage[available,functional]{usenixbadges/usenixbadges}
\graphicspath{{usenixbadges/}}
\newcommand{\yu}[1]{\textcolor[rgb]{0, 0, 0}{#1}} 
\definecolor{shepherdblue}{RGB}{0,82,170}

\begin{document}
\pagestyle{empty}
\title{\Sys: Data-Balanced Heterogeneous Pipeline Parallelism for Training Large Models
}

\date{}
\author{
  \begin{tabular}{c}
    {\rm \href{mailto:mianjie.yu@connect.um.edu.mo}{\textcolor{black}{Mianjie Yu}}\textsuperscript{1}, Zhongmin Zhao\textsuperscript{1},
    Zizhao Mo\textsuperscript{1,2,*}, Jianxiong Liao\textsuperscript{3},
    Wenxuan Li\textsuperscript{1}, Yulin Qiao\textsuperscript{1}}\\
    {\rm \href{mailto:huanlexu@um.edu.mo}{\textcolor{black}{Huanle Xu}}\textsuperscript{1,\textdagger}, Wenquan Yang\textsuperscript{4},
    Yongqiang Yang\textsuperscript{4}, Zeren Li\textsuperscript{4}, Ruifeng Tang\textsuperscript{4},
    \href{mailto:czxu@um.edu.mo}{\textcolor{black}{Chengzhong Xu}}\textsuperscript{1,\textdagger}}\\[.3em]\hspace{-.8em}
    \textsuperscript{1}University of Macau \quad
    \textsuperscript{2}Great Bay University \quad
    \textsuperscript{3}Sun Yat-sen University \quad
    \textsuperscript{4}Independent Researchers
  \end{tabular}
} %

\maketitle
\begingroup
\renewcommand{\thefootnote}{*}
\footnotetext{Work done while Zizhao Mo was at the University of Macau.}
\endgroup
\begingroup
\renewcommand{\thefootnote}{\textdagger}
\footnotetext{Corresponding authors: Huanle Xu and Chengzhong Xu.}
\endgroup

\begin{abstract}
Current heterogeneity-aware training systems for large models balance workloads by assigning unequal numbers of layers across pipeline stages—a strategy aimed at equalizing execution time through a unique pipeline configuration. However, this approach fails to address critical memory constraints and suffers from inherently coarse balancing granularity.

In this paper, we propose Heterogeneous Pipeline Parallelism (HPP), a novel approach enabling efficient and fine-grained data balance across heterogeneous pipelines. To address HPP's key challenges—prohibitive search overhead for optimal configurations and high communication overhead from asymmetric All-Reduce—we design the ETH training system. ETH employs a hierarchical workload balance framework that jointly optimizes balance across heterogeneous pipelines and within individual pipeline stages. To mitigate asymmetric communication costs, ETH implements block-wise gradient aggregation. Furthermore, it integrates specialized optimizations that dramatically reduce HPP configuration search overhead. Evaluations demonstrate ETH achieves up to 2$\times$ training speedup while reducing configuration generation overhead by up to 56$\times$ compared to state-of-the-art systems. The code is available
at \href{https://github.com/YuMJie/ETH}{https://github.com/YuMJie/ETH}.

\end{abstract}

\vspace{-.4em}
\section{Introduction}
Training large-scale models~\cite{brown2020language,grok3-cost,guo2025deepseek,comanici2025gemini,claude4,llama4,yang2025qwen3} on massive datasets requires extensive parallelism across numerous accelerators. Selecting optimal parallelism configurations is critical, as distribution strategies significantly impact training efficiency. Meanwhile, accelerator hardware advances rapidly but at increasing cost~\cite{100k-gpu-cluster}, creating economic and practical challenges: the exclusive use of cutting-edge accelerators leaves older hardware underutilized, inflating costs and reducing resource efficiency. Supply constraints further limit the availability of the latest accelerators for large-scale training~\cite{nvidia-backlog}. Consequently, effectively leveraging heterogeneous accelerators has become essential for scalable and cost-efficient large model training.

\begin{figure}
    \centering    
    \includegraphics[width=1\linewidth]{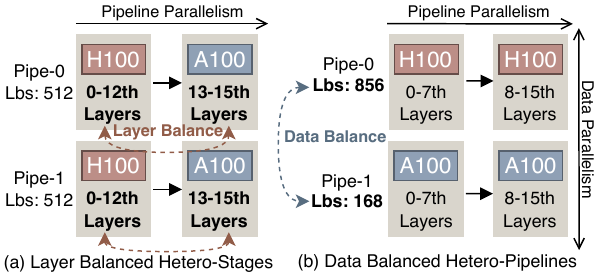}\vspace{-.8em}
    \caption{Layer-balancing versus data-balancing.}
\vspace{-1.3em}
\label{fig:balance}
\end{figure}

Several heterogeneity-aware training systems have been proposed, including SDPipe~\cite{miao2023sdpipe}, AMP~\cite{li2022amp}, Asteroid~\cite{ye2024asteroid}, HexiScale~\cite{yan2026hexiscale}, and Metis~\cite{um2024metis}. These systems primarily balance workloads through pipeline parallelism (PP) by assigning different numbers of layers to stages while maintaining unique pipeline configurations as demonstrated in Fig.~\ref{fig:balance}(a). Although this approach permits stages to employ distinct parallelization strategies such as tensor parallelism (TP) or data parallelism (DP), our analysis reveals significant device underutilization in heterogeneous clusters. Specifically, computation latency varies substantially across devices—with the slowest device taking up to 1.92$\times$ longer than others (detailed in \cref{sec:limitation}). This imbalance severely degrades training efficiency, as the synchronized nature of distributed training forces all devices to wait for the slowest one to complete computation.

Our analysis identifies the root cause of device underutilization: workload balancing via layer partitioning inherently struggles with memory and granularity constraints, making an optimal balance across heterogeneous devices difficult to achieve. 
First, implementing layer balancing—where in more computational layers are allocated to high-power devices to optimize throughput—inevitably increases memory overhead. Ideally, layer balancing achieves optimal resource utilization when the disparity in computational capability across heterogeneous devices aligns with the disparity in their memory capacities. However, recent hardware trends exhibit a pronounced imbalance between the growth of compute and memory: over the past five years, computation power has increased by approximately 16$\times$, whereas memory capacity has grown by less than 5$\times$. This divergence results in a fundamental mismatch between compute capability and memory capacity in modern accelerators. For instance, Ascend 910B2 roughly 1.2$\times$ higher computational throughput than 910B3 while offering the same memory capacity. As a consequence, conventional layer-balancing strategies fail to fully utilize the computational potential of high-performed but memory-constrained devices, leaving up to \textbf{32\%} of their computation power idle, as demonstrated in \cref{sec:limitation}.
Second, the limited number of model layers leads to inherently coarse-grained workload distribution across devices. For example, training GPT3-13B model requires approximately 64 devices using 8-way pipeline parallelism (PP), where each pipeline stage is assigned only 5 layers. Under such a configuration, adding or removing a single layer changes the per-device workload by 20\%. As a result, this coarse-grained layer assignment makes it difficult to achieve fine-grained workload balance, introducing an additional \textbf{16\%} pipeline bubbles in PP (\cref{sec:limitation}).

\begin{figure}
    \centering    
    \includegraphics[width=1\linewidth]{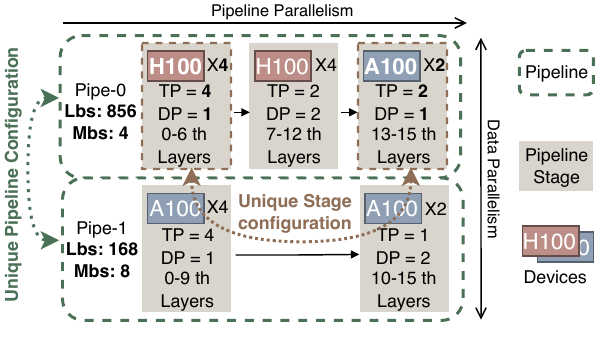}\vspace{-1em}
    \caption{An example of HPP configuration. HPP not only supports the training configuration with independent pipelines that have different local batch size, micro-batch size, but also supports the unique stage configurations that have different devices, layers and parallelism strategies. }
\label{fig:example}
\vspace{-1.2em}
\end{figure}

Fortunately, we observe that data balancing is more effective than layer balancing for harnessing heterogeneous devices. As illustrated in Fig.~\ref{fig:balance}, by exploiting the large global batch size commonly used in training, data balancing partitions the global batch into uneven local batches across pipelines to achieve finer-grained workload assignment, thereby effectively improving the resource utilization. This approach eliminates the need to align memory capacity with computational throughput across heterogeneous devices, since memory consumption is determined solely by the micro-batch size within each pipeline. To realize this idea, we propose a novel heterogeneous training paradigm—data-balanced heterogeneous pipeline parallelism (HPP), \yu{which exposes a broader space of independently configured
  heterogeneous pipelines and local-batch allocations.} This paradigm employs multiple independent, heterogeneous pipelines for data parallel training, with each pipeline assigned distinct local batch sizes to balance workloads and equalize the ending time, as shown in Fig.~\ref{fig:example}. Within each pipeline, different stages may adopt distinct DP/TP configurations, while workload balance is achieved through appropriate layer assignment.

  However, efficiently training large-scale models using HPP presents significant challenges. First, HPP enables independent pipelines to adopt distinct parallelization strategies, introducing a new optimization dimension beyond existing
  training systems. The combinatorial explosion of heterogeneous-pipeline configurations, coupled with the complexity of balancing stages within each pipeline, yields a \textbf{extremely huge  search space}. Exploring this space for optimal
  configurations incurs prohibitively long optimization times, even exceeding model training duration. Second, HPP introduces \textbf{asymmetric All-Reduce requirements}, which increase communication overhead. Specifically, devices sharing
  the same model parameters must perform gradient All-Reduce to ensure correct parameter updates. However, these gradients may reside at different memory addresses and have mismatched tensor sizes across devices due to heterogeneous pipeline
  configurations. Resolving these discrepancies requires additional padding and multi-stream synchronization to ensure address and size consistency, creating a bottleneck that significantly degrades communication performance as illustrated
  in \cref{HPP-challenges}.

This paper proposes \Sys, a heterogeneity-aware system built upon HPP with high efficiency on heterogeneous clusters. \Sys incorporates a joint workload balancing framework that optimizes configurations at two levels: dynamically adjusting local batch sizes across pipelines while rebalancing stage workloads within each pipeline using these updated batch sizes to derive optimal configurations. To minimize communication overhead, \Sys implements block-wise gradient aggregation, eliminating address/size inconsistencies through gradient partitioning and combining, while merging gradients within communication groups for optimized stream scheduling. For configuration search acceleration, \Sys employs pipeline caching to avoid redundant computations, throughput prediction to early skip searching low-performance device groups, priority-based search ordering to evaluate high-throughput candidate device groups first, and dynamic programming to reduce device group combinations.

We implement \Sys by extending Megatron-LM~\cite{shoeybi2019megatron}—and conduct comprehensive experiments across diverse accelerators and scales. Our evaluation shows that \Sys achieves up to 2$\times$ faster training and reduces the configuration generation time by up to 56$\times$ compared to exisiting training systems.

\vspace{-.4em}
\section{Background and Motivation}
\subsection{Distributing Training Parallelism}

\textbf{Pipeline parallelism (PP).}
PP partitions model layers into sequential stages—each comprising consecutive layers—and assigns these stages to accelerator groups. Existing PP methods optimize accelerator utilization by processing micro-batches in pipeline ~\cite{huang2019gpipe,narayanan2019pipedream,narayanan2021memory}. This approach splits minibatches into multiple micro-batches whose execution is orchestrated through the pipeline stages. {Larger micro-batch size brings higher throughput but incurs higher memory overhead.}

\noindent \textbf{Data parallelism (DP).} DP allows each accelerator to execute identical model layers on disjoint subsets of the input data~\cite{dean2012large,zinkevich2010parallelized}. The size of each subset, referred to as the local batch size, influences execution time but not memory consumption, as it determines the number of micro-batches, whereas memory usage is mainly governed by the micro-batch size. 

\noindent \textbf{Tensor parallelism (TP).} TP provides a complementary strategy for computational scaling within accelerator groups. Unlike DP's data-level partitioning, TP splits layer weights and computations into non-overlapping tensors~\cite{shoeybi2019megatron,narayanan2021efficient}, distributing these across a TP group. Each accelerator processes assigned tensors for every split layer while performing synchronous communication with other group members.

\begin{figure}[t]
\centering
\includegraphics[width=\linewidth]{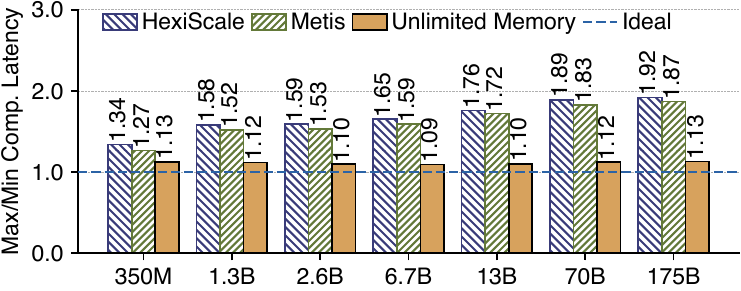}
\vspace{-2em}
\caption{The latency gap under different layer balancing methods across various models. We compare maximum-to-minimum device execution latency ratios for four approaches: a) HexiScale's layer-balancing (\textit{HexiScale}), b) Metis' layer-balancing (\textit{Metis}), c) Metis without memory constraints (\textit{Unlimited Memory}), and d) ideal balance (\textit{Ideal}). }
\label{fig:layer_balance_rate}
\vspace{-1.5em}
\end{figure}

\begin{figure*}[th]
\centering    \vspace{-.1em}
\begin{minipage}{0.33\textwidth}
\centering\includegraphics[width=1\linewidth]{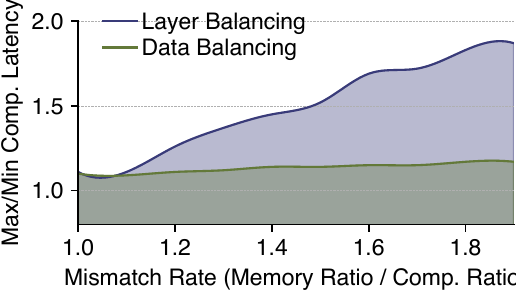}\vspace{-.5em}
\subcaption{The latency gap of layer balancing and data balancing parallelism approaches}
\label{fig:compare_balance_rate}
\end{minipage}
\begin{minipage}{0.33\textwidth}
\centering
\includegraphics[width=1\linewidth]{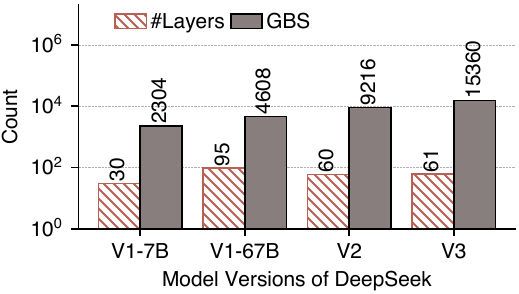}\vspace{-.5em}
\subcaption{The number of layers and global batch size settings across different models}
\label{fig:layer_vs_gbs_GPT}
\end{minipage}
\begin{minipage}{0.33\textwidth}
\centering\includegraphics[width=1\linewidth]{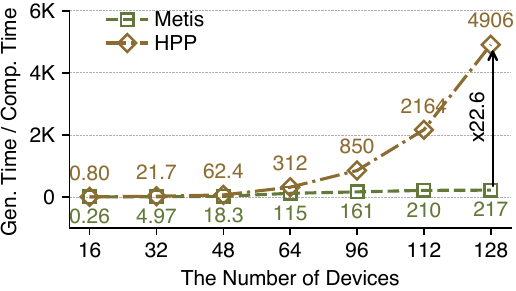}\vspace{-.5em}
\subcaption{Configuration generation overhead across different scales (GPT3-13B)}  
\label{fig:search_vs_execute_metis_our}
\end{minipage}
\vspace{-.8em}
\caption{The advantage of data balance and the high search overhead for HPP.}  
\vspace{-1.2em}
\label{motivation-balancing}
\end{figure*}

\subsection{Limitations of Existing Works}\label{sec:limitation}
  While several systems~\cite{miao2023sdpipe,ye2024asteroid,li2022amp,um2024metis,yan2026hexiscale} address heterogeneous training, their workload balancing strategy primarily relies on varying the number of layers assigned to
  pipeline stages to equalize stage execution times. Notably, Metis~\cite{um2024metis} supports distinct stage parallelization strategies—DP, TP, or hybrids—but still balances execution time by assigning different layer counts across stages.
  Although HexiScale~\cite{yan2026hexiscale} can search over multiple pipelines, it focuses solely on balancing workload within each pipeline by adjusting layer assignments. Consequently, the balancing mechanism across these
  approaches relies on layer balancing: distributing layers unevenly across pipeline stages to balanced execution times.

However, this balancing mechanism faces fundamental limitations in both granularity and memory impact: (1) The inherently limited number of layers constrains balancing precision due to coarse granularity; (2) more critically, layer allocation dictates device memory consumption (parameters, gradients, optimizer states), which can trigger recomputation when exceeding hardware limits—paradoxically exacerbating imbalance. 
Our experiments on Ascend 910B2/910B3 heterogeneous clusters (model configurations in Tab.~\ref{tab:model_config}) illustrate these constraints through Fig.~\ref{fig:layer_balance_rate}, which compares maximum-to-minimum device execution latency ratios for four approaches: 
{(a) HexiScale's layer-balancing (\textit{HexiScale}),
(b) Metis' layer-balancing (\textit{Metis}), and to demonstrate the impact of memory on layer balance, we add (c) Metis without memory constraints (\textit{Unlimited Memory}) and add (d) ideal balance (\textit{Ideal}) to demonstrate the coarse granularity of layer balance.}
Two key trends emerge. First, latency ratios for HexiScale and Metis layer-balancing increase significantly with model size—rising from 1.34 and 1.27 at 350M to 1.92 and 1.87 at 175B, respectively, causing in up to 32\% computation power idle—while the unlimited-memory ratio remains stable between 1.09 and 1.13. This shows that memory bottlenecks progressively undermine balance efficacy as models scale. Second, even without memory constraints, a persistent gap from ideal balance remains: \yu{unlimited-memory latency ratios exceed the ideal ratio by 9--13\%, resulting in 16\% additional pipeline bubbles.} This gap is attributed to the coarse granularity of layer-based strategies, which cannot eliminate inherent latency disparities across pipeline stages.

\subsection{Opportunities}
  \yu{Data balancing across heterogeneous pipelines significantly outperforms existing layer-balancing methods.} This presents a strong opportunity to accelerate LLM training.

Data balancing improves workload balance in heterogeneous training. To illustrate this advantage, we conduct a controlled study on heterogeneous B2/B3 devices by varying the memory-compute mismatch rate through the available memory of B2, defined as the memory ratio divided by the compute ratio between heterogeneous devices. In this study, the global batch size is set equal to the number of layers. We compare data balancing with Metis' layer-balancing strategy using the maximum-to-minimum computation latency ratio in GPT3-13B, where a value closer to 1 indicates better balance. As shown in Fig.~\ref{fig:compare_balance_rate}, layer balancing degrades markedly as the mismatch rate increases, with the latency ratio rising from 1.11 to 1.87 as the mismatch rate grows from 1.0 to 1.9. In contrast, data balancing remains stable (1.09 to 1.17) across all settings. These results show that data balancing remains effective even under severe memory-compute mismatch and largely removes the need to align memory capacity with computational throughput across heterogeneous devices.

Moreover, the effectiveness of data balancing scales with global batch size, as larger values enable finer workload granularity. To demonstrate data balancing's applicability in practical training scenarios, we analyze global batch configurations from DeepSeek~\cite{bi2024deepseek,liu2024deepseekv2,liu2024deepseekv3}. As shown in Fig.~\ref{fig:layer_vs_gbs_GPT}, global batch sizes substantially exceed layer counts. In the most extreme case (DeepSeekV3), the batch size exceeds layer count by $\geq$251$\times$. Even the smallest disparity (V1-67B) maintains a $\geq$48$\times$ ratio. This exceptionally fine granularity enables data balancing to achieve higher load distribution efficiency than layer balancing. Consequently, by configuring different local and micro-batch sizes across pipelines, data balancing supports more flexible workload balancing.

\vspace{-.4em}
\section{Data-Balanced Heterogeneous Pipeline Parallelism and Challenges}\label{HPP-challenges}
Motivated by this observation, we propose a novel heterogeneous training paradigm—data-balanced heterogeneous pipeline parallelism (HPP).  
Our paradigm employs multiple independent heterogeneous pipelines for data parallel training and supports flexible training strategies.
As in the HPP example shown in Fig.~\ref{fig:example},
within each pipeline, different stages can use different DP/TP while balancing workloads through layer counts. 
For each pipeline, HPP uses distinct local batch sizes to balance workloads between pipelines by equaling ending time.
HPP enables effective data balancing across heterogeneous pipelines, significantly outperforming existing layer-balancing methods. This presents a strong opportunity to accelerate LLM training, but requires solving new challenges in pipeline generation and communication mechanisms:

\textbf{Challenge 1: The vast search space for an optimal data-balanced HPP.} Exhaustive search over the HPP configuration space is prohibitively expensive. 
Unlike conventional heterogeneous parallelism—which optimizes layer balancing within a unique pipeline—HPP requires configuring heterogeneous independent pipelines with distinct parallelization strategies, 
drastically expanding the search space. 
We quantify search overhead relative to iteration time for both Metis's layer balancing and data-balanced HPP running GPT3-13B training on 910B2/B4 heterogeneous clusters. 
As Fig.~\ref{fig:search_vs_execute_metis_our} demonstrates,
the search costs of the optimal HPP configuration substantially exceed the training overhead reduction achieved through HPP. 
Moreover, this imbalance escalates with scale: as the number of devices increases, configuration generation time grows superlinearly relative to training time.
For data-balanced HPP, the generation-to-computation ratio ranges from 0.8 to 4905—representing a 3-22$\times$ increase over layer-balancing schemes.
Overall, this prohibitive overhead critically undermines HPP's scalability in practice.

\textbf{Challenge 2: High communication overhead of asymmetric All-Reduce.}
In data parallelism, devices replicate a portion of the model parameters. We refer to devices sharing duplicated parameters as a DP group. Before updating model parameters, the devices of each DP group synchronizes gradients via All-Reduce collective operation. In layer-balanced configurations, this All-Reduce is straightforward because all devices in the DP group hold identical parameters. However, in data-balanced HPP systems—where devices use different parallelism strategies across heterogeneous pipelines—devices of the same DP group may store non-identical parameters. Consequently, the address and size of the gradients requiring synchronization become inconsistent. 

\begin{figure}
    \centering
    \includegraphics[width=1\linewidth]{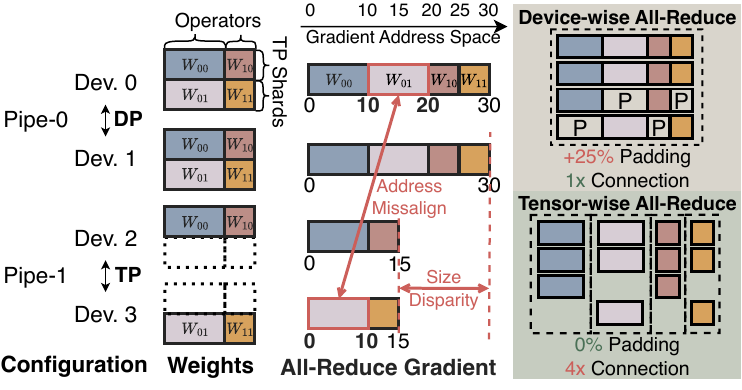}
    \vspace{-2.1em}
\caption{Asymmetric All-Reduce across four devices. Device-wise All-Reduce pads the missing gradient to ensure consistency, but brings the padding overhead. Tensor-wise All-Reduce splits the gradient by operator tensor.
    }   
\vspace{-1.6em} \label{fig:Asymmetric All-Reduce}
\end{figure}

\begin{figure*}
    \centering
    \includegraphics[width=1\linewidth]{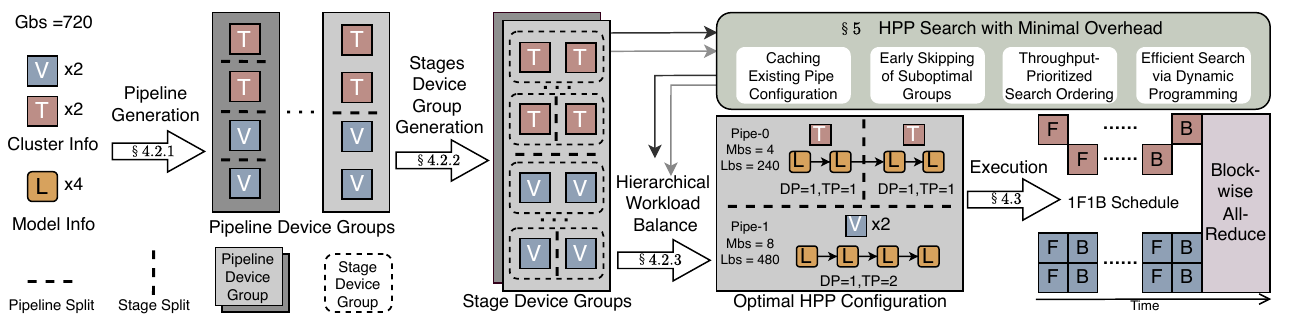}
    \vspace{-2em}
   \caption{Overview of \Sys system.} 
\vspace{-1.5em}
    \label{fig:overview}
\end{figure*}

As shown in Fig.~\ref{fig:Asymmetric All-Reduce}, we demonstrate a two-pipeline configuration. The first pipeline (Devices 0 and 1) employs DP, while the second pipeline (Devices 2 and 3) uses TP. Despite sharing duplicated weights requiring gradient synchronization via All-Reduce, direct synchronization is prevented by two inconsistencies: 1) \textbf{Address misalignment}: TP partitioning in Devices 2 and 3 results in partial weight maintenance. This causes gradient addressing mismatches—for example, layer 0's first operator gradients (Light purple) occupy addresses 0-10 on Device 3 versus 10-20 on Device 0’s memory space.
2) \textbf{Size disparity}: All-Reduce operations require uniform gradient dimensions, yet DP devices (Devices 0 and 1) produce 30-element gradients while TP devices (Devices 2 and 3) generate 15-element gradients.
The device-wise padding approach is the simplest solution for gradient inconsistencies, ensuring alignment 
and size uniformity through missing gradient padding. However, this introduces significant communication and memory overhead from redundant data. While tensor-wise All-Reduce reduces memory overhead by splitting gradients at smallest tensor granularity, it incurs excessive communication overhead by establishing separate All-Reduce connections per tensor—creating inefficient multi-stream bottlenecks. This trade-off necessitates a coarse-grained synchronization approach that optimizes asymmetric communication to reduce HPP overhead while enabling large-scale training.

\vspace{-.4em}
\section{System Design}
\vspace{-.3em}
This section presents the system architecture of \Sys, a framework specifically designed to address the critical challenges of HPP in both of generation and communication.

 \vspace{-.5em}
\subsection{System Overview}
\label{sec:System Overview}

Fig.~\ref{fig:overview} illustrates \Sys's complete architecture and workflow. The process involves generating heterogeneous pipelines (\cref{sec:Devices mapping Pipeline Groups}), creating stage-device groups (\cref{sec:Devices mapping Stage Groups}), and hierarchical workload balancing (\cref{sec:Hierarchical workload balance}). After selecting configurations, \Sys executes pipelines independently. Upon backward pass completion, \Sys performs gradient synchronization via block-wise All-Reduce (\cref{sec:Runtime Optimization}).
\Sys addresses high generation costs through several optimizations that reduce search overhead (\cref{sec:Pipeline-Level Pruning}), including: configuration caching, early skipping of suboptimal groups, throughput-prioritized search ordering, and dynamic programming.

\subsection{Configuration of Heterogeneous Pipelines}
\label{sec:Hetero-Aware Search Space}

\subsubsection{Pipeline generation}\label{sec:Devices mapping Pipeline Groups}

\Sys generates all possible pipeline device group configurations for heterogeneous accelerators, ensuring each configuration is unique in terms of execution time. The process works as follows: \Sys first iterates over the possible number of pipelines, ranging from 1 (all devices in one pipeline) to $N$ (each device in its own separate pipeline). For each pipeline count, \Sys generates distinct groupings of devices. Crucially, pipelines are considered unordered sets; only the composition matters, not their sequence. Therefore, configurations like $[\langle V, V \rangle, \langle T, T \rangle]$ and $[\langle T, T \rangle, \langle V, V \rangle]$ represent the same pipeline device group.
\textbf{Example}: Given two $V$-type and two $T$-type accelerators, \Sys generates configurations ranging from a single pipeline $[\langle V, V, T, T \rangle]$ to the maximum split $[\langle V \rangle, \langle V \rangle, \langle T \rangle, \langle T \rangle]$, along with intermediate groupings like $[\langle V, V \rangle, \langle T, T \rangle]$ (but only once, due to unordered pipelines).

\subsubsection{Stage device groups}
\label{sec:Devices mapping Stage Groups}
After generating pipeline device groups, \Sys creates stage device groups for each pipeline within a pipeline device group. For a pipeline represented as $\langle V, V, T, T\rangle$, \Sys generates stage device groups ranging from one stage$[\{V, V, T, T\}]$ to four stages $[\{V\}, \{V\}, \{T\}, \{T\}]$, with $[,]$ denoting a stage device group and $\{,\}$ denoting an individual stage; crucially, stage ordering matters as $[\{V, V\}, \{T, T\}]$ and $[\{T, T\}, \{V, V\}]$ represent distinct partitions: the former places both $V$-type accelerators in the first stage and $T$-types in the second, while the latter reverses this arrangement. Since activation overhead is higher in earlier stages under the 1F1B pipeline schedule~\cite{narayanan2019pipedream,fan2021dapple}, \Sys distinguishes these ordered configurations.

\vspace{-.5em}
\subsubsection{Hierarchical workload balance}\label{sec:Hierarchical workload balance}
{After generating stage device groups, \Sys searches for efficient pipeline configurations and balances workloads across pipelines. The search selects each stage's TP/DP degrees, layer count, and recomputation operations, together with each pipeline's micro-batch and local batch sizes. These choices are interdependent: changing a pipeline's local batch size can change its preferred stage configuration and execution time.}

{\Sys uses a three-phase heuristic to address this dependency. It first optimizes stage configurations with each pipeline's local batch size initialized to the global batch size, then allocates local batches across pipelines, and finally refines stage configurations using the assigned local batches. This procedure seeks low predicted iteration time; it does not guarantee a globally optimal joint configuration.}

{\textbf{Stage configuration search.} During stage workload balancing, \Sys seeks to reduce pipeline latency} for a given global batch size ($\text{Gbs}$) by determining the degree of TP ($tp$), DP ($dp$), the number of layers ($l$), recomputation operators ($ro$) for each stage and micro-batch size ($\text{MBS}$). 
\Sys's stage-specific recomputation reduces memory overhead through finer granularity.
 The total latency combines: 1) the summed stage execution times $\sum_{j} T_{j}$ for the first micro-batch, 
 and 2) the slowest stage's latency $\max_{j} T_{j}$ multiplied by remaining micro-batches ($\text{Gbs}/\text{Mbs} - 1$). 
 The stage-level optimization objective is:
\vspace{-.5em}
\begin{equation}
\mathop{\min}\limits_{\{tp,dp,ro,l,\text{Mbs}\}} 
\left[ 
    \sum_{j} T_{j} + \max_{j} T_{j} \cdot \left(\frac{\text{Gbs}}{\text{Mbs}} - 1\right)
\right],
\label{eq:layer_balance} \vspace{-.4em}
\end{equation}
where $T_j$ denotes the latency of $j$-th stage .
\begin{figure*}[!h]
    \centering
    \includegraphics[width=\linewidth]{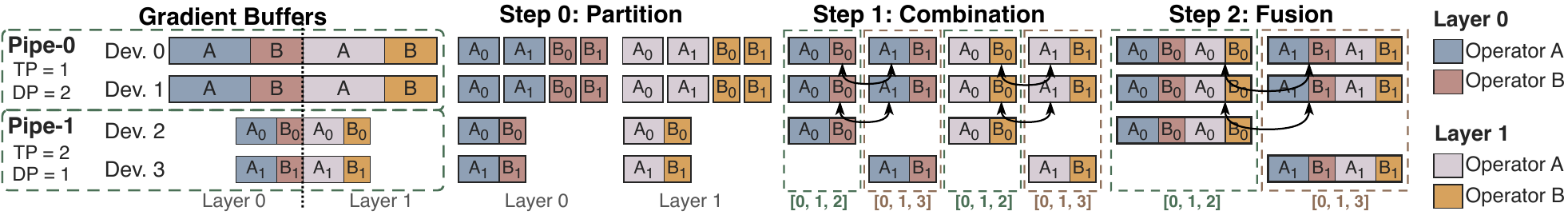}
    \vspace{-1.8em}
    \caption{The workflow of Block-wise All-Reduce. \Sys ensures address consistency and size consistency through partition and combination, then reduce the connections by merging the gradient block which belong to the same communication group.}
\label{fig:All_reduce_compare}
\vspace{-1.2em}
\end{figure*}
{\Sys builds on the stage-optimization primitives of Aceso~\cite{liu2024aceso}, which searches parallel configurations for homogeneous clusters. Aceso uses latency and memory estimates to identify stage bottlenecks, then iteratively adjusts operator placement, devices count, micro-batch size, and stage-specific recomputation to alleviate these bottlenecks until performance no longer improve.}

{\Sys adapts these operations to heterogeneous stage device groups. For a latency bottleneck, it moves operators to a less-loaded stage, increases DP while decreasing TP, or increases the micro-batch size. For a memory bottleneck, it moves operators to a lower-memory stage, increases TP while decreasing DP, reduces the micro-batch size, or enables recomputation.}

\textbf{Local batch allocation.} With stage configurations initialized, \Sys assigns distinct local batch sizes to balance workloads across pipelines. Consider a device group comprising $P$ pipelines, where the $i$-th pipeline employs micro-batch size $\text{Mbs}_i$. The objective is to choose local batch sizes $\text{Lbs}_i$ that minimize the maximum completion time across all stages and pipelines.
The optimization problem is formulated as follows:
\vspace{-1.2em}
\begin{align}
    \mathop{\min}\limits_{\{\text{Lbs}_i\}}
    &
    \left[
    \max\limits_{i \in[1,P]} 
    \left( 
    \max\limits_{j \in [1,S_i]} Te_{i}^{j}
    \right) 
    \right]
    \label{eq:data_balance} \\
    \text{subject to} \quad 
    & \sum_{i=1}^P \text{Lbs}_i \geq \text{Gbs}
    \label{cond:gbs}, \\ 
    & \text{Lbs}_i \bmod \text{Mbs}_i = 0, \quad  \forall i \in [1,P].
    \label{cond:smaple}
    \vspace{-.4em}
\end{align}
Here $S_i$ denotes the number of stages in the $i$-th pipeline.
The time $Te_{i}^{j}$ for stage $j$ in pipeline $i$ combines four components: the summed forward times for the first micro-batch ($\sum_{k=1}^{S_i} Tf_{i}^{k}$), the slowest stage latency scaled by remaining micro-batches ($\max_{k} T_{i}^{k} \cdot (\text{Lbs}_i/\text{Mbs}_i - 1)$), the summed backward times from stage $j$ to the final stage for the last micro-batch ($\sum_{k=j}^{S_i} Tb_{i}^{k}$), and the All-Reduce overhead ($AR_{i}^{j}$):
\vspace{-.4em}
\begin{equation}
    Te_{i}^{j} = 
    \sum_{k=1}^{S_i} Tf_{i}^{k} + 
    \max_{k \in [1,S_i]} T_{i}^{k} \cdot \left( \frac{\text{Lbs}_i}{\text{Mbs}_i} - 1 \right) + 
    \sum_{k=j}^{S_i} Tb_{i}^{k} + 
    AR_{i}^{j}.
    \vspace{-.2em}
\end{equation}
The per-micro-batch timings $Tf_{i}^{k}$ (forward), $Tb_{i}^{k}$ (backward), and $T_{i}^{k}$ (combined forward-backward) for stage $k$ in pipeline $i$ are obtained through system profiling. The All-Reduce overhead $AR_{i}^{j}$ depends on gradient size $\text{Gra}_{i}^{j}$, number of communicating devices $N_{i}^{j}$, and available bandwidth $BW_{i}^{j}$:
\vspace{-.4em}
\begin{equation}
    AR_{i}^{j} = \frac{2 \cdot \text{Gra}_{i}^{j} \cdot (N_{i}^{j} - 1)}{N_{i}^{j} \cdot BW_{i}^{j}}.
    \vspace{-.3em}
\end{equation}
This formulation captures communication costs during gradient synchronization between pipeline stages.

\Sys uses a greedy allocation procedure for the objective in Eq.~\eqref{eq:data_balance}. The algorithm initializes $\text{Lbs}_i = 0$ for all pipelines. It then proceeds iteratively: In each iteration, \Sys identifies the pipeline where incrementing $\text{Lbs}_i$ by $\text{Mbs}_i$ (satisfying Constraint~\eqref{cond:smaple}) would yield the smallest increase in the maximum end time across all pipelines. The selected pipeline's local batch size is increased by its micro-batch size, while other pipelines' batch sizes remain unchanged. This process repeats until the aggregate local batch size satisfies Constraint~\eqref{cond:gbs}. \par
{\textbf{Stage configuration refinement.} Changing local batch sizes also changes the stage-level optimization objective. The initial stage configurations derived in Eq.~\eqref{eq:layer_balance} were optimized for the global batch size rather than the newly assigned local batch sizes. To address this dependency, \Sys refines each pipeline's stage configuration by substituting the global batch size in Eq.~\eqref{eq:layer_balance} with its corresponding local batch size $\text{Lbs}_i$.

\subsection{Execution of Heterogeneous Pipelines}\label{sec:Runtime Optimization}

As highlighted in \cref{HPP-challenges}, the HPP mechanism of \Sys introduces two inconsistencies during gradient synchronization, which poses significant challenges for All-Reduce. Device-wise solutions incur extra memory and communication overhead due to additional padding, whereas tensor-wise All-Reduce requires a separate connection for each tensor shard on each device. This design creates inefficient multi-stream bottlenecks, especially because each GPT-3 layer contains 13 operators.

To address this challenge, \Sys introduces block-wise All-Reduce, a method that reduces memory and communication overhead by operating at the block granularity. As illustrated in Fig.~\ref{fig:All_reduce_compare}, \Sys resolves the two inconsistencies through \textbf{gradient partitioning} and \textbf{combination}, ensuring all devices maintain block consistency with the device possessing the maximum TP degree (e.g., Devices 2 and 3 with $\text{TP}=2$). \Sys further employs block fusion to reduce the total number of blocks.
During the initialization phase, each device establishes DP communication groups with devices sharing duplicated parameters across pipeline stages.
Within each DP group, devices subsequently exchange information about their layer index range, rank in TP group and TP degree ($tp$). Utilizing the above information, each device identifies both the layer/TP block assignments of peer devices and the maximum TP degree ($tp_{\max}$) within its DP communication group. \Sys then executes gradient partitioning and combination to enforce block-wise consistency.

\textbf{Partitioning:}
All devices first partition gradients layer by layer. Devices with a TP degree lower than $tp_{\max}$ further partition their gradients along the operator and TP-shard dimensions, producing $tp_{\max}/(\text{local TP degree})$ partitions for each operator. For example, Devices 0 \& 1 with lower TP degrees, split their contiguous blue operator blocks into finer-grained partitions, yielding behavior equivalent to tensor-wise All-Reduce. In contrast, Devices 2 and 3, which operate at $tp_{\max}$, partition gradients only once along the layer dimension.

\textbf{Combination:}
During combination, blocks with lower TP degrees are  aligned with those at the highest TP degree, while blocks already at the highest TP degree require no further operation. For example, on Devices 0 and 1, which have lower TP degrees, \Sys first exchanges the $A_1$ and $B_0$ gradient blocks. It then merges the $A_0$ and $B_0$ gradients to match the state of Device 2, which has the highest TP degree. The combination stage ensures uniform All-Reduce block sizes and consistent memory addressing across all devices. Moreover, based on the layer assignments and TP blocks of each device, \Sys appends DP communication group information to the combined gradient blocks for subsequent All-Reduce communication.

\textbf{Fusion:}
Although partitioning and combination guarantee gradient consistency, we observe redundant DP communication groups in certain blocks (e.g., blocks 0 and 2 share identical DP group: Devices 0, 1, 2). To optimize, each device detects blocks that share the same DP  group and merges them in layer order, further reducing 
All-Reduce operations.

\section{HPP Search with Minimal Overhead}
\label{sec:Pipeline-Level Pruning}
{Searching the HPP configuration space is expensive for two reasons: stage optimization repeatedly solves identical pipeline subproblems, and the number of candidate pipeline combinations grows rapidly with cluster size. \Sys addresses these costs through four complementary optimizations. Caching avoids repeated optimization of identical pipeline subproblems. Early skipping avoids fully evaluating device groups whose estimated throughput bounds fall below the best configuration found so far. Search ordering evaluates promising groups first, improving the effectiveness of early skipping. Finally, dynamic programming reuses overlapping partition results to reduce the cost of exploring pipeline combinations.}

\subsection{Caching Existing Pipeline Configurations}\label{sec:Cache for Pipeline configuration}
We observe that identical pipeline segments recur across generated device groups. For example, with two $V$-type and two $T$-type accelerators, \Sys generates configurations like $[\langle V, V\rangle, \langle T, T\rangle]$ and $[\langle V, V\rangle, \langle T\rangle, \langle T\rangle]$—both containing the $\langle V, V\rangle$ pipeline. \Sys caches the best configurations found (including TP/DP degrees, layer count, recomputation operations per stage, and micro-batch size) during a pipeline's first optimization. Subsequent occurrences of $\langle V, V\rangle$ in any device group then reuse this cached configuration directly.

\subsection{Search via Early Skipping}\label{sec:Prediction-based pruning}
Building on pipeline caching, \Sys implements efficient pruning to eliminate unpromising configurations early in the search process. For each candidate pipeline device group, \Sys computes a throughput upper bound for each pipeline using both deterministic and approximate methods, using the smaller value as the pruning estimate. If this total bound of all pipelines falls below the throughput of the best configuration found so far, \Sys immediately skips further evaluation of the candidate pipeline device group.

The deterministic upper bound is derived as follows: if every device in pipeline $P_1$ has a unique corresponding device in pipeline $P_2$ with strictly greater memory capacity and computation power, then $\text{throughput}(P_2)$ constitutes an upper bound for $\text{throughput}(P_1)$. To efficiently verify this relationship, \Sys employs a bipartite graph algorithm. Given device sets $\mathcal{D}_1$ for $P_1$ and $\mathcal{D}_2$ for $P_2$ where $|\mathcal{D}_1| \leq |\mathcal{D}_2|$, we construct edges between $d_i \in \mathcal{D}_1$ and $d_j \in \mathcal{D}_2$ if and only if:\vspace{-.3em}
\[ 
\text{mem}(d_j) > \text{mem}(d_i) \land \text{comp}(d_j) > \text{comp}(d_i). \vspace{-.3em}
\] 
After building this bipartite graph, \Sys applies the Hungarian algorithm to find the maximum matching. When the matching size equals $|\mathcal{D}_1|$, $\text{throughput}(P_2)$ is confirmed as an upper bound for $\text{throughput}(P_1)$.

\Sys estimates the approximate throughput upper bound for the candidate pipeline by benchmarking against similar pipelines stored in its cache. This approach leverages two critical hardware characteristics that impact pipeline performance: Higher computational capacity reduces computation time and directly increases throughput, while greater memory capacity enables both larger micro-batch sizes and reduced operator recomputation requirements, collectively improving pipeline efficiency. \Sys estimates current pipeline's throughput by multiplicative scaling: 
The cached pipeline's throughput is scaled by both the computation power ratio (current to cached) and memory capacity ratio (current to cached), yielding what we term the \textit{approximate throughput}. To accommodate potential estimation errors, this value is then multiplied by a configurable fault tolerance factor $\alpha \geq 1$, producing the final \textit{approximate upper bound} metric.

\subsection{Optimization for the Search Order}
\label{sec:Search Order Optimization}
Early identification of high-throughput configurations is essential for \Sys's efficiency, enabling aggressive skipping of suboptimal candidates during the search process.
Our analysis reveals a key correlation: Pipeline device groups exhibiting lower computation power variance—both between pipelines ($V_{inter}$) and within pipelines ($V_{intra}$)—demonstrate higher probability of achieving superior throughput. 

Formally, let $Com_{i}^j$ denote the computation power of the $j$-th device in the $i$-th pipeline. \Sys computes the inter-pipeline variance $V_{inter}$ as:
\vspace{-1em}
\begin{equation}
    V_{inter} = \mathop{\mathrm{Var}}\limits_{i \in [1,P]}\left( \sum_{j=1}^{N_i} Com_{i}^j \right),
    \vspace{-.3em}
\end{equation}
where $N_i$ represents the device count in pipeline $i$ and $P$ is the total pipeline count. The intra-pipeline variance $V_{intra}$ aggregates per-pipeline device variances:
\vspace{-.8em}
\begin{equation}
    V_{intra} = \sum_{i=1}^{P} \mathop{\mathrm{Var}}\limits_{j \in [1,N_i]}\left( Com_{i}^j \right).
    \vspace{-.7em}
\end{equation}
To validate this relationship, we sampled the best configurations found for 100 randomly generated pipeline device groups across varying global batch sizes. Experiments are conducted on 910B2/B4 devices running GPT3-13B. Configurations are sorted by ascending variance values, with corresponding throughput measurements plotted in Fig.~\ref{fig:varp_vars_B2B4_GBS}. The results confirm an inverse correlation: Throughput decreases as $V_{inter}$ increases.
Furthermore, increasing the global batch size (Gbs) enhances configuration throughput, an effect particularly pronounced in configurations with high inter-pipeline variance ($V_{inter}$). Similarly, lower intra-pipeline variance ($V_{intra}$) correlates strongly with improved throughput, as evidenced in Fig.~\ref{fig:varp_vars_B2B4_LAYER}. However, this relationship exhibits layer-dependent attenuation: As model depth ($L$) increases, the impact of $V_{intra}$ on throughput progressively diminishes.

Motivated by these insights, we design a heuristic cost function to prioritize high-throughput pipeline device groups during configuration search. The function incorporates two normalized variance components:
\vspace{-.7em}
\begin{equation}
\text{Cost} = \frac{V_{inter}}{\text{Gbs}} + \frac{V_{intra}}{L},
\vspace{-.3em}
\end{equation}
where $\text{Gbs}$ denotes global batch size and $L$ denotes the number of model layers. This formulation inherently favors configurations with: (1) balanced computation across pipelines and (2) homogeneous devices within pipelines.
Finally, we can obtain the cost value of each pipeline device group, giving priority to the group with the lowest cost value among groups with same pipeline count to generate and evaluate.

\begin{figure}
 \centering
\begin{minipage}{0.235\textwidth}
\centering
\includegraphics[width=1\linewidth]{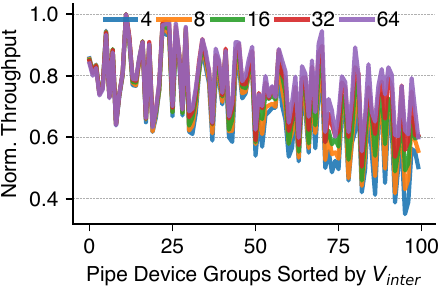} \vspace{-1.5em} \subcaption{Throughput of optimal configuration across $V_{inter}$ values at different global batch sizes}  
\label{fig:varp_vars_B2B4_GBS}
\end{minipage}
\begin{minipage}{0.235\textwidth}
\center\includegraphics[width=1\linewidth]{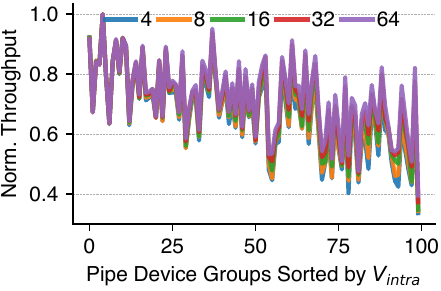}
\vspace{-1.5em}
\subcaption{Throughput of optimal configuration across $V_{intra}$ values at different layer counts}  \label{fig:varp_vars_B2B4_LAYER}
\end{minipage}
\vspace{-.7em}
\caption{Throughput sensitivity of heterogeneous pipeline configurations to $V_{inter}$ and $V_{intra}$.}
\vspace{-1.3em}
\label{fig:varp}
\end{figure}

\vspace{-.2em}
\subsection{Search via Dynamic Programming} \label{sec:Dynamic Programming Optimization}
In large-scale clusters with thousands of devices, techniques like pipeline caching and early skipping remain ineffective at substantially reducing search overhead. This limitation stems from the initial search overhead being primarily dominated by hierarchical workload balancing. Although pipeline caching and early skipping can optimize this balancing phase, they become insufficient as the search progresses. Once most viable pipeline configurations are cached, the fundamental bottleneck shifts to an exponential explosion in possible pipeline combinations. To address this combinatorial complexity at scale,  \Sys also incorporates a dynamic programming approach to search for efficient HPP configurations.

\Sys initiates the search for efficient HPP configurations by processing the number of pipelines within each pipeline device group in ascending order. Given a cluster of $n$ devices and a predetermined pipeline device group order (detailed in \cref{sec:Search Order Optimization}), the configuration $R_p^n$ with the highest estimated throughput for a candidate group with $p$ pipelines is determined by identifying a split point $i \in [1, n-1]$ that maximizes throughput. This optimization partitions the cluster so that the first $i$ devices are distributed across $p-1$ pipelines ($R_{p-1}^i$), while devices $i+1$ through $n$ form a single pipeline ($\mathrm{pipe}_{i+1}^n$). The throughput transition equation of the dynamic programming formulation is:
\vspace{-.4em}
\begin{equation}
\mathrm{Tp}(R_p^n) = \max_{i \in [1, n-1]} 
\left[ 
    \mathrm{Tp}(R_{p-1}^i) + \mathrm{Tp}(\mathrm{pipe}_{i+1}^n)
\right],
\vspace{-.6em}
\end{equation}
where $\mathrm{Tp}(\cdot)$ denotes throughput of a pipeline configuration.

After obtaining the HPP configuration via dynamic programming, \Sys balances pipelines in configuration using data-balancing and refines each pipeline stage's configuration for this HPP setup (\cref{sec:Hierarchical workload balance}). Notably, \Sys continuously adjusts device arrangements across pipelines and triggers the dynamic programming search whenever early skipping efficiency becomes too low, continuing until all pipeline device groups are covered. For each fixed device ordering, the dynamic programming recurrence has a complexity of $O(PN^2)$, where $P$ is the maximum number of pipelines and $N$ is the total number of devices. As $P$ is typically equal to $N$, the complexity is $O(N^3)$. Thus, \Sys reduces the problem's complexity from exponential to polynomial time.

\section{System Implementation}

We implemented \Sys by adding approximately 9,000 lines of Python code to Megatron-LM~\cite{shoeybi2019megatron}. Our modifications include rewriting the generator to leverage Aceso's stage-specific operation recomputation to expand the search space, dynamically allocating different balancing times to stage device groups to control per-stage search cost, and enabling ZeRO-1 and ZeRO-2 for optimizer-state and gradient sharding~\cite{rajbhandari2020zero}.

\noindent\textbf{Runtime and memory profiling.}
  \yu{Before configuration search, \Sys profiles the target model once on each accelerator type. For every candidate micro-batch size and TP degree, the profiler executes each operator type in
  the target model and records its forward and backward latency. It also benchmarks the collective and point-to-point communication operations over the message sizes and device links
  considered by the search. Under the same settings, the profiler records the parameter and activation sizes of each operator and the memory reserved by PyTorch.
    }

\noindent\textbf{Adaptive balance time allocation}. Unlike Aceso, which limits convergence time in stage workload balancing with a fixed time budget and returns the best configuration found once the budget is exhausted, \Sys adopts an adaptive strategy. \Sys allocates balancing time per stage device group based on the group's inter-stage memory and computation variances and those of peer groups (with identical stage counts). The rationale is that higher variance leads to suboptimal performance, thus deserving less time budget. Specifically, \Sys normalizes each group's variances against peer groups and adjusts its time budget to be divided by the product of the normalized memory and computation variances.

 \begin{table}[t]
    \centering
         \vspace{-.5em}
        \caption{The configurations for heterogeneous accelerators}
         \vspace{-.5em}
         \resizebox{\linewidth}{!}{%

    \begin{tabular}{c|c|c|c}
    \hline
        Clusters & Comp. Power & Memory  &  Bandwidth  \\  \hline
         Ascend  910B2 & 376 TFLOPS & 64 GB& 1536 Gbps\\  
         Ascend  910B3 & 313 TFLOPS  & 64 GB & 1536 Gbps  \\  
         Ascend  910B4 & 280 TFLOPS & 32 GB & 736 Gbps  \\
         NVIDIA A100 & 312 TFLOPS & 80 GB & 1536 Gbps  \\
         NVIDIA RTX3090 & 80 TFLOPS & 24 GB & 1408 Gbps  \\
          \hline
    \end{tabular}}
    \vspace{-1.1em}
    \label{tab:device_tflops}
\end{table}
\noindent\textbf{Stage-specific operation recomputation}.
Recomputation saves device memory by discarding activations after forward passes and recalculating them during backward passes. Megatron-LM supports only coarse-grained strategies: full recomputation, which deletes all activations, or
  selective recomputation~\cite{korthikanti2023reducing}, which excludes operations such as dropout. By contrast, \Sys  decomposes the LLM into operators (e.g., QKV projections and matrix multiplications) for fine-grained, stage-specific recomputation. If a stage's
  estimated memory exceeds device capacity during planning, \Sys prioritizes recomputing high-memory, low-recompute-cost operations until the stage fits. The HPP configuration records this plan; during training, \Sys discards
  only the selected activations after the forward pass and recomputes them just in time for gradient calculations, enabling precise memory control.

\vspace{-.2em}
\section{System Evaluation}
\subsection{Experimental Setup}
\textbf{Cluster environments}.
We conduct experiments on two hardware platforms: Atlas cluster with Ascend 910B2, 910B3, and 910B4 NPUs, and an NVIDIA GPU cluster with A100 and RTX 3090. These devices differ in compute capability, memory capacity, and communication performance, with detailed specifications summarized in Tab.~\ref{tab:device_tflops}.
Each node hosts up to 8 devices, providing 200~Gbps intra-node bandwidth for homogeneous configurations and 100~Gbps inter-node bandwidth for heterogeneous configurations. 

\noindent\textbf{Workloads}.
To demonstrate \Sys's capability in rapidly finding optimized configurations across model scales, we evaluate it on GPT-3~\cite{brown2020language} (350M to 175B parameters) and LLaMA-2 70B~\cite{touvron2023LLaMA} trained across heterogeneous clusters of varying sizes. {Referring to Metis}, model configurations and corresponding device scales are detailed in Tab.~\ref{tab:model_config}. For main evaluations, we vary accelerators per node (2--8) and nodes per cluster (2--128) across different model sizes, using a sequence length of 2048 and global batch size of 1024.

\noindent\textbf{Baselines}.
We compare \Sys wtih the SOTA heterogeneity-aware training systems HexiScale~\cite{yan2026hexiscale} and Metis~\cite{um2024metis}.

\vspace{-.5em}
\subsection{End-to-End Performance}
\noindent\textbf{Training performance.}
Fig.~\ref{fig:execute_search_B234_1024_hete} (Left) shows per-iteration training times. \Sys consistently outperforms prior systems across varying model sizes. 
Compared to HexiScale, \Sys achieves speedups of 1.36-1.55$\times$, 1.42-2.00$\times$, and 1.63-1.82$\times$ on the $B2$/$B3$, $B2$/$B4$, and $B3$/$B4$ heterogeneous clusters, respectively. 
This stems from \Sys's heterogeneous pipeline configurations and enlarged search space enabling better workload distribution. 
\Sys also achieves 1.20-1.84$\times$ speedups over Metis, which struggles with effective data balance, whereas \Sys leverages large global batch sizes. 
As device computational gaps grow, \Sys's advantage over Metis increases because wider gaps worsen load imbalances. 
For example, \Sys achieves 1.32-1.84$\times$ speedups on $B2$/$B4$ and 1.22-1.64$\times$ on $B3$/$B4$, compared to 1.20-1.40$\times$ on $B2$/$B3$. This is because $B2$ and $B3$ have smaller compute power and memory capacity gaps than either has with $B4$. 

\begin{table}[t]
    \centering
      \vspace{-.5em}
    \caption{The various models and corresponding cluster scale. $N\times M$ represents $N$ nodes with $M$ devices for each node, $A_i$ and $A_j$ present different accelerators}
    \vspace{-.5em}
    \resizebox{\linewidth}{!}{%

    \begin{tabular}{c|c|c|c} \hline
         Model & \#Layer & Hidden Dim    & Clusters Scales \\ \hline
         350M & 24  & 1024   & 1$\times$2$\times$$A_i$ + 1$\times$2$\times$$A_j$ \\
         1.3B & 24  & 2048   & 2$\times$2$\times$$A_i$ + 2$\times$2$\times$$A_j$\\
         2.6B & 32  & 2560   & 2$\times$4$\times$$A_i$ + 2$\times$4$\times$$A_j$\\
         6.7B & 32  & 4096   & 4$\times$4$\times$$A_i$ + 4$\times$4$\times$$A_j$\\
         13B & 40  & 5120   & 4$\times$8$\times$$A_i$ + 4$\times$8$\times$$A_j$\\ 
         70B & 80  & 8192   & 32$\times$8$\times$$A_i$ + 32$\times$8$\times$$A_j$\\ 
         175B & 96  & 12288   & 64$\times$8$\times$$A_i$ + 64$\times$8$\times$$A_j$\\ \hline
    \end{tabular}}
    \vspace{-1.1em}
    \label{tab:model_config}
\end{table}

\begin{figure}[t]
\vspace{-.1em}
    \centering
    \includegraphics[width=\linewidth]{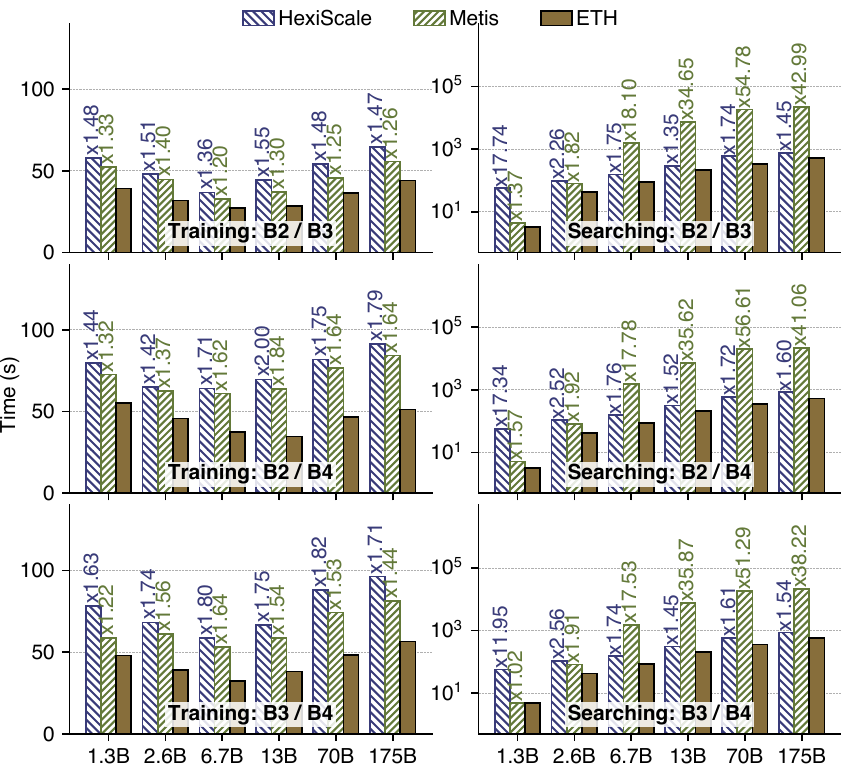}
    \vspace{-1.5em}
    \caption{Average training time per iteration and generation time in different models and heterogeneous clusters.}
    \label{fig:execute_search_B234_1024_hete}
    \vspace{-1.2em}
\end{figure}

\noindent\textbf{Search overhead.}
Since configuration search time impacts system availability, we compare search times for all systems across models and heterogeneous clusters. As shown in Fig.~\ref{fig:execute_search_B234_1024_hete} (Right), 
\Sys reduces search time by 1.35-17.74$\times$ ($B2$/$B3$), 1.52-17.34$\times$ ($B2$/$B4$), and 1.45-11.95$\times$ ($B3$/$B4$) versus HexiScale. This stems from \Sys's pipeline cache optimization that eliminates redundant searches—unlike HexiScale's two-phase approach.
\Sys reduces configuration generation time by 38.22--42.99$\times$ over Metis for 175B models through its search optimizations.
\Sys thus demonstrates superior search efficiency compared to baselines in most scenarios.

\noindent\textbf{Visualization of Configuration.}
\yu{We visualize the parallelization strategies for GPT-3 13B on $B2$/$B4$ clusters.  Due to space constraints, the detailed results are deferred to Appendix~\ref{sec:appendix_visualization}.}

\subsection{Computation and Communication Time}
We experiment on $B2$/$B4$ clusters to analyze the breakdown of iteration time across different methods and evaluate the communication time of various All-Reduce mechanisms.

\noindent\textbf{Computation time.}
We compare slowest-stage training times across model sizes for \Sys, HexiScale, and Metis. As shown in Fig.~\ref{fig:time_break_down_B2_B4}, we analyze computation time, pipeline bubbles, and communication time. \Sys outperforms both
  baselines in computation. For 13B, \Sys computes 30\% faster than Metis and 40\% faster than HexiScale. \Sys uses DP between and within pipelines, accelerating computation relative to TP. Regarding pipeline bubbles, \Sys reduces the 13B
  model's bubble size to 60\% of HexiScale's and 75\% of Metis's. The baselines' layer balancing causes significant stage imbalances, whereas \Sys reduces bubbles by evenly distributing workloads across pipelines through data balancing. \Sys
  incurs slightly higher communication overhead because HPP employs more DP, requiring All-Reduce communication across heterogeneous devices at the end of each iteration.

\noindent\textbf{Block-wise All-Reduce.}
  To evaluate \Sys's block-wise All-Reduce, we compare it with device-wise and tensor-wise All-Reduce across models. As shown in Fig.~\ref{fig:allreduce_vs_compute}, block-wise All-Reduce achieves 1.38-1.71$\times$ and 1.56-2.74$\times$
  communication speedups over tensor-wise and device-wise All-Reduce, respectively. Its advantage over device-wise All-Reduce grows with model size because the latter pads missing tensors, causing significant padding overhead. Block-wise
  All-Reduce also outperforms tensor-wise All-Reduce, whose finer granularity creates dozens or hundreds of communication streams per device, severely hurting bandwidth utilization. In contrast, block-wise All-Reduce transmits
  data in larger blocks, significantly reducing the number of communication streams.

  \begin{figure}[t]
    \centering
        \begin{minipage}{0.235\textwidth}
    \centering
    \includegraphics[width=1\linewidth]{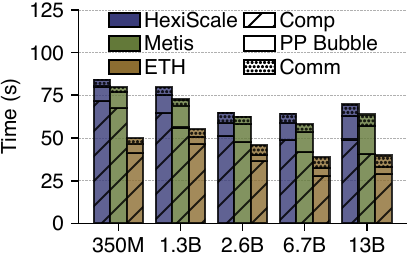}
    \vspace{-1.5em}
    \subcaption{Breakdown of iteration time}
    \label{fig:time_break_down_B2_B4}
    \end{minipage}
    \begin{minipage}{0.235\textwidth}
    \centering
    \includegraphics[width=1\linewidth]{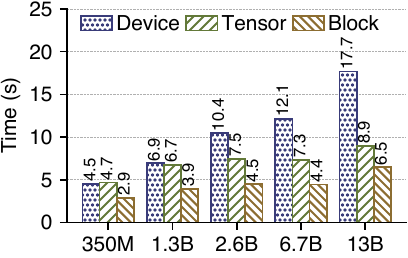}
    \vspace{-1.5em}
    \subcaption{Latency of All-Reduce}
    \label{fig:allreduce_vs_compute}
    \end{minipage}
    \vspace{-.9em}
\caption{The performance of three All-Reduce mechanisms and the breakdown of iteration time of different methods in various models on $B2$/$B4$ heterogeneous clusters.}
\label{fig:allreduce_vs_compute_time_break_down_B2_B4} 
\vspace{-1.2em}
\end{figure}

\subsection{System Scalability}
To evaluate the scalability of each system, we conduct experiments in 13B on $B2$/$B4$ heterogeneous clusters, varying cluster size, global batch size, and sequence length. {Moreover, we also evaluat \Sys on NVIDIA heterogeneous clusters.}

\noindent\textbf{Scalability w.r.t. \#devices.}
We evaluate \Sys on clusters using 8 devices per node. As show in Fig.~\ref{fig:ablation_scale_13B_combined}, \Sys achieves up to 1.86$\times$ and 1.83$\times$ speedups over HexiScale and Metis, respectively. \Sys's advantage over Metis scales efficiently, growing from a 1.25$\times$ speedup at 16 devices to 1.77$\times$ at 128 devices. While HexiScale encounters OOM at 16, 32, and 48 devices, \Sys trains successfully across all scales. For searching time, \Sys is 1.47$\times$ to 2.48$\times$ faster than HexiScale, and 1.59$\times$ to 35.61$\times$ faster than Metis. Specifically, \Sys accelerates generation times in large clusters, outperforming Metis by 1.59$\times$ at 16 devices and up to 14$\times$ at 128 devices.

\noindent\textbf{Scalability w.r.t. global batch size.}
To evaluate performance across global batch sizes, we experiment with sizes from $2^2$ to $2^{12}$. As shown in Fig.~\ref{fig:ablation_gbs_13B_combined}, \Sys achieves 1.10$\times$ to 2.00$\times$ and 1.02$\times$ to 1.83$\times$
  execution-time speedups over HexiScale and Metis. \Sys's performance advantage grows with global batch size. When the global batch size is 4, \Sys speeds up 1.02$\times$ to 1.10$\times$ compared to baselines. As the global batch size
  increases to the largest, \Sys achieves speedups of 1.64$\times$ and 1.44$\times$. This is because \Sys uses heterogeneous pipelines with data balance; large global batch sizes bring finer granularity for workload balancing. Although
  \Sys uses global batch size for load balancing, search time remains stable (204.62--217.49\,s) and consistently below the baselines.

\begin{figure*}[t]
    \centering
    \begin{minipage}{0.24\textwidth}
    \centering
    \includegraphics[width=1\linewidth]{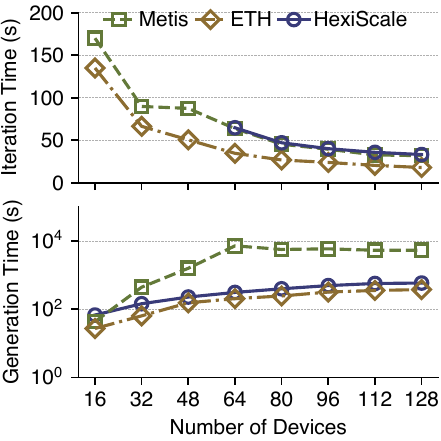}
    \vspace{-1.5em}
    \subcaption{\yu{Number of devices}}
    \label{fig:ablation_scale_13B_combined}
    \end{minipage}
    \begin{minipage}{0.24\textwidth}
    \centering
    \includegraphics[width=1\linewidth]{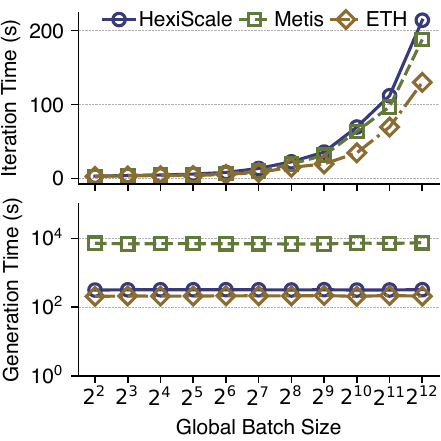}
        \vspace{-1.5em}
    \subcaption{\yu{Global batch size}}
    \label{fig:ablation_gbs_13B_combined}
    \end{minipage}
    \begin{minipage}{0.24\textwidth}
    \centering
    \includegraphics[width=1\linewidth]{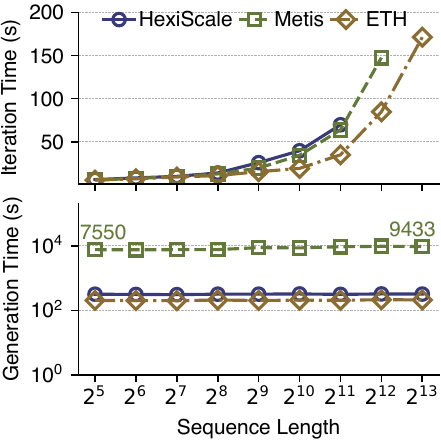}
    \vspace{-1.5em}
    \subcaption{\yu{Sequence length}}
    \label{fig:ablation_seq_13B_combined}
    \end{minipage}
    \begin{minipage}{0.24\textwidth}
    \centering
    \includegraphics[width=1\linewidth]{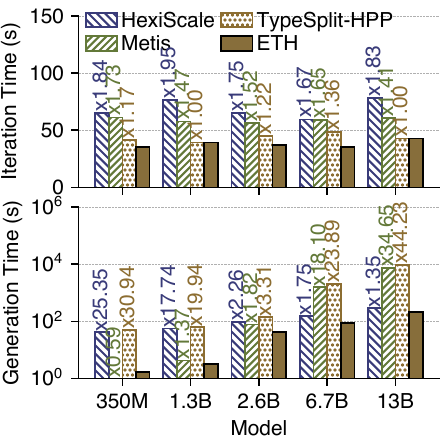}
    \vspace{-1.5em}
    \subcaption{\yu{NVIDIA GPU evaluation}}
    \label{fig:type_split_combined}
    \end{minipage}
    \vspace{-.8em}
    \caption{\yu{Scalability and global-search evaluation. (a)-(c) Performance comparison on GPT-3 13B when varying device count, GBS, and sequence length on $B2$/$B4$ clusters. (d) TypeSplit-HPP comparison across GPT-3 models on A100/RTX 3090 clusters.}}
    \vspace{-1.em}
    \label{fig:scalability_type_split}
\end{figure*}

\noindent\textbf{Scalability w.r.t. sequence length.}
To evaluate performance, we conduct experiments with sequence lengths ranging from $2^5$ to $2^{13}$. As shown in Fig.~\ref{fig:ablation_seq_13B_combined}, \Sys achieves maximum speedups of 2.08$\times$ and 1.83$\times$ over HexiScale and Metis, respectively. As sequence length increases, \Sys's speedup grows because larger per-layer activation sizes make baselines prone to memory insufficiency during layer balancing. Furthermore, HexiScale and Metis support maximum sequence lengths of 2048 and 4096, respectively, while \Sys reaches 8192. The same panel shows that sequence length minimally impacts \Sys's and HexiScale's search costs, whereas Metis's overhead varies. For example, search costs for \Sys, HexiScale, and Metis range from 200.49\,s to 214.34\,s, 307.28\,s to 323.23\,s, and 7550.25\,s to 9612.77\,s, respectively.
 
\noindent\textbf{Scalability on NVIDIA GPUs.}
We further demonstrate the scalability of \Sys on NVIDIA GPUs, consisting of A100 and RTX 3090 GPUs (100 Gbps interconnect, matching the setup in Tab.~\ref{tab:model_config}). As shown in Fig.~\ref{fig:type_split_combined}, \Sys delivers excellent performance across GPT-3 models ranging from 350M to 13B parameters.
In terms of training time, \Sys reduces iteration times by 1.41–1.73$\times$ compared to Metis and by 1.67–1.95$\times$ compared to HexiScale. The improvements in generation time are even more significant, with speedups of up to 34.65$\times$ and 25.35$\times$ over Metis and HexiScale, respectively.

\subsection{\yu{Evaluation of Key Components}}

\noindent\textbf{\yu{Global configuration search.}}
\yu{To test whether global search is necessary, we design TypeSplit-HPP strategy. It partitions devices by type and independently optimizes each homogeneous pipeline with \Sys's stage optimizer.
 As shown in Fig.~\ref{fig:type_split_combined}, \Sys matches TypeSplit-HPP on two models and achieves 1.17--1.36$\times$ speedups on the other three. Type splitting is therefore sufficient for two models but misses faster configurations for the other three. Despite its larger search space, \Sys reduces search cost by 3.31--44.23$\times$, making full HPP practical when it improves configuration quality.}

\noindent\textbf{\yu{Memory-saving strategies.}}
\yu{To isolate the direct effect of \Sys's memory-saving strategy (stage-specific recomputation), we disable it while keeping device assignment, parallelism degrees, batch sizes, and workload unchanged. As shown in Fig.~\ref{fig:memory_A100RXT3090_1024_hete}. peak memory only increases by 1.8\%--5.7\% across the evaluated model configurations.}

\noindent\textbf{\yu{Predictor accuracy.}}
\yu{As shown in Fig.~\ref{fig:predictor_GPU}, runtime and memory prediction errors range from $-6.7\%$ to $+6.6\%$ and from $+4.9\%$ to $+7.4\%$, respectively. Both remain within 7.4\%, supporting candidate evaluation, while consistent memory overestimation avoids understating memory requirements.}

\subsection{The Effectiveness of Search Optimization}

\noindent\textbf{Ablation study on search optimization.}
To demonstrate the impact of our various search optimizations on search cost, Fig.~\ref{fig:search_ablation_combined} illustrates the results of \Sys from no optimizations up to all optimizations, by adding a new optimization each time based on the previous experiment. 
For search cost, unoptimized \Sys is 9844\,s, exceeding HexiScale and Metis by over 20$\times$ and 1.3$\times$, respectively. Moreover, \textit{SES} shows a 1.16$\times$ speedup over the previous experiment. When combined with \textit{OSO}, \Sys facilitates earlier searches for better configurations, thereby assisting \textit{SES} enhance by 2.09$\times$. In this ablation, the optimizations preserve the iteration time of the generated configuration while reducing search cost. This is an empirical result for the evaluated workload, rather than a general guarantee that pruning preserves the global optimum.

\noindent\textbf{Parameter sensitivity.}
  $\alpha$ is a hyperparameter that controls the size of the approximate upper bound for early skipping. We vary $\alpha$ from 0.5 to 4 to evaluate its impact on configuration-generation time and
  configuration quality. As shown in Fig.~\ref{fig:ablation_alpha_combined}, when $\alpha$ is 0.5, the search cost is the smallest at 127.64\,s, but the iteration time of the generated configuration is the largest at 38.14\,s. That is
  because early skipping depends on a smaller approximate upper bound, which can prune higher-throughput configurations.
  When $\alpha$ reaches more than 1.5, its impact on upper-bound pruning weakens because early skipping depends on a smaller deterministic upper bound, leading to near-constant search cost and training time.
  Additionally, the results indicate that $\alpha$ only needs to be greater than 1. For instance, at $\alpha$ equal to 4, the search time is about 238\,s, only 10\% longer than when $\alpha$ equals 1.5.

\begin{figure}[t]
    \begin{minipage}{0.234\textwidth}
    \centering
    \includegraphics[width=1\linewidth]{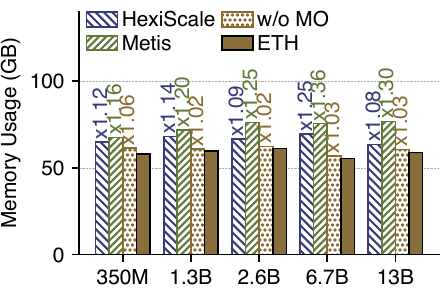}
    \vspace{-1.5em}
    \subcaption{\yu{Peak memory}}
    \label{fig:memory_A100RXT3090_1024_hete}
    \end{minipage}
    \begin{minipage}{0.236\textwidth}
    \centering
    \includegraphics[width=1\linewidth]{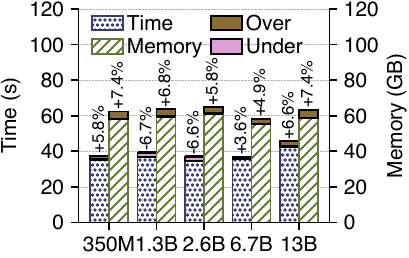}
    \vspace{-1.5em}
    \subcaption{\yu{Predictor accuracy}}
    \label{fig:predictor_GPU}
    \end{minipage}
    \vspace{-.8em}
    \caption{\yu{Memory efficiency and predictor accuracy across GPT-3 models on A100/RTX 3090 clusters. \textit{w/o MO} disables \Sys's memory-saving strategies; positive and negative errors denote overestimation and underestimation, respectively.}}
    \vspace{-1.em}
    \label{fig:memory_predictor_GPU}
\end{figure}

\begin{figure}[t]
    \begin{minipage}{0.235\textwidth}
    \centering
    \includegraphics[width=1\linewidth]{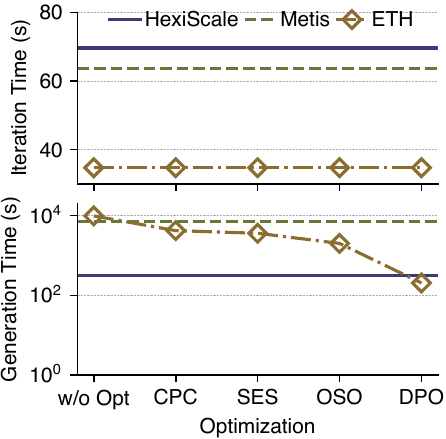}
    \vspace{-1.5em}
    \subcaption{\yu{Search-optimization ablation}}
    \label{fig:search_ablation_combined}
    \end{minipage}
    \begin{minipage}{0.235\textwidth}
    \centering
    \includegraphics[width=1\linewidth]{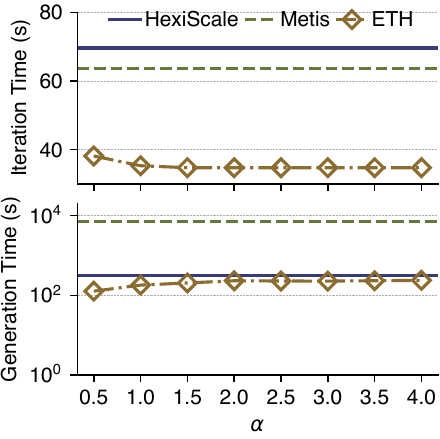}
    \vspace{-1.5em}
    \subcaption{\yu{$\alpha$ sensitivity}}
    \label{fig:ablation_alpha_combined}
    \end{minipage}
    \vspace{-.8em}
    \caption{\yu{Search-optimization evaluation for GPT-3 13B on $B2$/$B4$ clusters: (a) cumulative ablation and (b) sensitivity to tolerance factor $\alpha$. \textit{CPC}: Caching Pipeline Configuration; \textit{SES}: Search via Early Skipping; \textit{OSO}: Optimization for Search Order; \textit{DPO}: Dynamic Programming Optimization.}}
    \vspace{-1.3em}
    \label{fig:search_optimization}
\end{figure}

\subsection{Model Convergence Performance}
We evaluate the convergence of \Sys against baselines using the LLaMA-2 70B model trained on the Enwiki dataset for 2,000 iterations. Experiments are conducted on $B2$ and $B3$ heterogeneous clusters, and to assess robustness, we evaluate \Sys on a $B3$ homogeneous cluster. 
As detailed in Appendix~\ref{App:Model Convergence Performance},
all methods exhibit nearly identical loss curves, with differences of approximately 0.1\%. Such minor discrepancies can be attributed to hardware precision and lie well within an acceptable margin of error. This variance arises because \Sys balances data distribution across pipelines without altering the data content, resulting in convergence behavior comparable to the baseline. Moreover, \Sys produces consistent loss curves across all clusters, demonstrating its robustness.

\vspace{-.4em}
\section{{Discussion}}
\label{sec:discussion}
\vspace{-.1em}
\noindent{\textbf{Search quality for HPP.} 
\Sys uses Aceso's~\cite{liu2024aceso} fast bottleneck-driven method to optimize stages within each pipeline.  Like gradient descent on a nonconvex objective, Aceso's local adjustments target current bottlenecks but may lead to local optima, so global optimality of the resulting HPP configuration is not guaranteed in \Sys.}

\noindent{\textbf{Additional parallelism dimensions.}
\Sys currently searches over PP, TP, DP. Incorporating context parallelism (CP)~\cite{liu2024ringattention,DEEPSPEEDULYSSES} or expert parallelism (EP)~\cite{lepikhin2021gshard} would introduce additional configuration variables and require models for context-partitioning communication or routing-dependent expert workloads. These extensions would also affect memory requirements and parameter placement, requiring corresponding changes to configuration search and gradient synchronization.}

\noindent{\textbf{Other pipeline schedules.}
\Sys's latency and memory models currently assume 1F1B~\cite{narayanan2019pipedream} scheduling. Other PP schedules~\cite{narayanan2021efficient,qi2024zero,wu2024bitpipebidirectionalinterleavedpipeline} may change pipeline bubbles and activation lifetimes, affecting both configuration feasibility and workload balance. Supporting them would require new cost models and validation of execution and synchronization order.}

\section{Related Works}
\textbf{Homogeneous Training Systems.}
% \vspace{-.2em}
Exisiting homogeneous training systems~\cite{zheng2022alpa,tarnawski2020efficient,mirhoseini2017device,jia2019beyond,wang2019supporting,jia2019beyond,eliad2021fine,zhong2025optimizing,jiang2024megascale,zhang2025disttrain} explore parallelism
and training techniques.
For PP, Mirhoseini et al.~\cite{mirhoseini2017device} use machine learning for device placement. For TP, FlexFlow~\cite{jia2019beyond} and Tofu~\cite{wang2019supporting} use cost models and dynamic programming to identify optimal
intra-operator parallelism strategies.
FTPipe~\cite{eliad2021fine} combines DP and PP for fine-tuning on commodity hardware. Piper~\cite{tarnawski2021piper} and Alpa~\cite{zheng2022alpa} propose hierarchical parallelism that incorporates
DP and TP at the lower level while applying PP at the upper level.
Unity~\cite{unger2022unity} integrates algebraic transformations with parallelization techniques. For recomputation, Adapipe~\cite{sun2024adapipe} adopts adaptive recomputation to determine optimal strategies, whereas
DTR~\cite{kirisame2020dynamic} introduces a greedy online algorithm for operator-level recomputation.

\noindent \textbf{Heterogeneity-aware Systems.}
Several works \cite{narayanan2020heterogeneity,chaudhary2020balancing,mo2026serving,mo2025hetis} address the execution of distributed DL on heterogeneous cluster. Whale~\cite{jia2022whale} balances computation and memory through device assignment, while Metis~\cite{um2024metis} and HexiScale~\cite{yan2026hexiscale} balance layers across PP stages. SDPipe~\cite{miao2023sdpipe} targets dynamic GPU performance
heterogeneity. HAP~\cite{zhang2024hap} introduces uneven workload distribution across both data and tensor dimensions. 
\yu{Cephalo~\cite{benson2024cephalo} assigns unequal batches under fixed FSDP~\cite{zhao2023pytorch}, whereas \Sys jointly optimizes local batch size and stage configurations.}
 NeutronHeter~\cite{10.1145/3749175} addresses compute, memory, and network heterogeneity through multi-level workload mapping. Cluster-level
  schedulers~\cite{gu2019tiresias,jayaram2023sia,mahajan2020themis,mo2024heet,mo2024optimal,qiao2021pollux,xiao2018gandiva,mo2025fast,weng2022mlaas} instead optimize jobs allocations in
  heterogeneous clusters.

% \vspace{-.4em}
\section{Conclusion}
% \vspace{-.1em}
This paper proposes HPP, a new training paradigm for large models in heterogeneous clusters.
Building upon HPP, we design \Sys, a training system that generates efficient HPP configurations with small overhead. \Sys incorporates a joint workload balancing framework operating at two levels: It dynamically adjusts local batch sizes across pipelines and simultaneously rebalances stage workloads within each pipeline using these updated batch sizes. Furthermore, \Sys 
implements a block-wise All-Reduce method to reduce communication overhead. The key contribution of \Sys is its ability to significantly expand the available configuration space for parallel training in heterogeneous environments with high scalability. 

\vspace{-.04em}
\section{Acknowledgment}
% \vspace{-.1em}
We thank our shepherd \href{https://stephanie-wang.github.io/}{\yu{Dr. Stephanie Wang}} and the anonymous NSDI reviewers for their valuable feedback. 
This work was supported in part by the Science and Technology Development Fund of Macau (0041/2025/RIA1, 0074/2025/AMJ), and the Multi-Year Research Grant of University of Macau (MYRG-GRG2024-00255-FST-UMDF, MYRG-GRG2025-00119-FST).

\clearpage
\appendix

\section{Visualization of Training Configuration}

\begin{table}[htbp]
\centering
\vspace{-.5em}
\caption{Visualization of parallelization strategy generated by Metis, HexiScale and \Sys.  \#L: number of layers in each stage. \textit{Qkv}: Query, key and value. \textit{Sco}: QK Score. \textit{AS}: Attention Softmax. \textit{AD}: Attention Dropout}\label{Tab:HPP_Visualization}
\resizebox{\linewidth}{!}{%

\begin{tabular}{|c|c|c|c|c|c|}
\hline
\multicolumn{6}{|c|}{\textbf{Metis}} \\ \hline
Pipeline & Stage & Dev. & TP/DP & \#L & Recomp. Operator \\
\hline
\multirowcell{8}{ Mbs=4 \\  Gbs=1024} & 0 & $B2\times8$ & 2/4 &5  &None  \\ 
\cline{2-6} & 1 & $B2\times8$ & 2/4 &6 &None  \\
\cline{2-6} & 2 & $B2\times8$ & 4/2 &7 &None  \\
\cline{2-6} & 3 & $B2\times8$ & 2/4 &6 &None \\
\cline{2-6} & 4 & $B4\times8$ & 8/1 &3 &None \\
\cline{2-6} & 5 & $B4\times8$ & 8/1 &4 &None \\
\cline{2-6} & 6 & $B4\times8$ & 4/2 &4 &None \\
\cline{2-6} & 7 & $B4\times8$ & 4/2 &5 &None \\
\hline
\hline

\multicolumn{6}{|c|}{\textbf{HexiScale}} \\ \hline
\multirowcell{5}{ Pipe-0 \\ Mbs=4 \\  Lbs=512} & 0,1,2 & $B4\times4$ & 4/1 & 3 &None  \\ 
\cline{2-6} & 3 & $B4\times4$ & 4/1 &4 &None \\
\cline{2-6} & 4 & $B2\times4$ & 4/1 &5 &None \\
\cline{2-6} & 5,6 & $B2\times4$ & 4/1 &7 &None \\
\cline{2-6} & 7 & $B2\times4$ & 4/1 &8 &None \\ \hline
Pipe-1 & \multicolumn{5}{c|}{Identical to Pipe-0}  \\ 
\hline\hline

\multicolumn{6}{|c|}{\textbf{\Sys}} \\
\hline
\multirowcell{5}{Pipe-0 \\ Mbs=4 \\  Lbs=344} & 0 & $B4\times4$ & 4/1 &4 &\textit{Qkv}, \textit{Sco}, 4\textit{AS}, 4\textit{AD}  \\ 
\cline{2-6} & 1 & $B4\times4$ & 4/1 &4 & 4\textit{AS}, 4\textit{AD} \\
\cline{2-6} & 2 & $B4\times4$ & 4/1 &5 &4\textit{AS}, 4\textit{AD} \\
\cline{2-6} & 3,4,5 & $B4\times4$ & 2/2 &5 &None \\
\cline{2-6} & 6,7 & $B4\times4$ & 1/4 &6 &None \\
\hline
\multirowcell{4}{Pipe-1 \\ Mbs=8 \\  Lbs=680} & 0 & $B2\times8$ & 8/1 &8 &6\textit{AS}, 6\textit{AD}	 \\ 
\cline{2-6} & 1 & $B2\times8$ & 8/1 &9 &4\textit{AS}, 4\textit{AD} \\
\cline{2-6} & 2 & $B2\times8$ & 4/2 &11 &None \\
\cline{2-6} & 3 & $B2\times8$ & 4/2 &12 &None \\ 
\hline
\end{tabular}}
\end{table}
\label{sec:appendix_visualization}

Table~\ref{Tab:HPP_Visualization} visualizes the parallelization strategies for GPT-3 13B on $B2$/$B4$ heterogeneous clusters. Metis uses a single pipeline and employs layer balancing to distribute workloads across stages, without utilizing data for load balancing. Consequently, Metis cannot fully exploit the global batch size. While HexiScale supports searching over multiple pipelines, it has two limitations: devices within a stage cannot perform data parallelism (DP), fixing the DP degree at 1, and pipelines are always identical since HexiScale cannot use data to balance workloads. Thus, HexiScale’s configuration reduces to a single pipeline with a DP degree of 2 per stage. In contrast, \Sys supports multiple heterogeneous pipelines and uses different local batch sizes to optimize workload distribution. In the example shown, \Sys utilizes two pipelines performing DP. The first pipeline ($B4$) processes 344 samples with 4 samples per micro-batch, while the second pipeline ($B2$) processes 680 samples with 8 samples per micro-batch. Furthermore, \Sys allows different stages to have unique configurations for DP/TP, layers, and recomputation. This demonstrates that \Sys possesses a significantly larger search space to identify efficient configurations, offering greater potential than state-of-the-art systems.

\section{Model Convergence Performance}
\label{App:Model Convergence Performance}

Figure~\ref{fig:Convergence} presents the loss curves of \Sys and the baselines on the LLaMA-2 70B model trained on the Enwiki dataset for 2,000 iterations. The experiments include $B2$/$B3$ heterogeneous clusters and a $B3$ homogeneous cluster for robustness evaluation. Across all settings, the loss curves are nearly identical, with differences of approximately 0.1\%. We also provide a zoomed-in view of the final iterations to highlight the small gap among different methods. These results further confirm that \Sys improves system efficiency without affecting model convergence.

\begin{figure}[t]
    \centering
    \includegraphics[width=\linewidth]{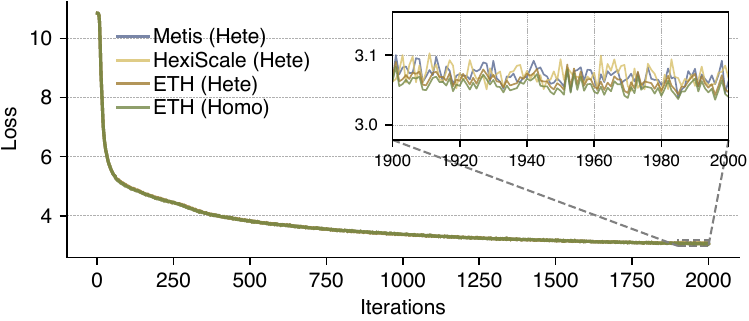}
    \caption{Loss curves of \Sys, HexiScale, and Metis on heterogeneous clusters, together with \Sys on a homogeneous cluster. The right panel zooms in on the final iterations.}
    \label{fig:Convergence}
\end{figure}
\end{document}